\documentstyle{article}
\newcommand   {\etal}    {{\it et~al.}}
\newcommand{\beq}{\begin{equation}}
\newcommand{\eeq}{\end{equation}}
\newcommand{\beqa}{\begin{eqnarray}}
\newcommand{\eeqa}{\end{eqnarray}}
\newcommand{\bea}{\begin{eqnarray}}
\newcommand{\eea}{\end{eqnarray}}
\newcommand   {\ev}[1]   {\langle #1\rangle}
\newcommand   {\rgyr}    {r_{\mbox{\tiny g}}}

\newcommand   {\ELJ}     {E_{\mbox{\tiny LJ}}}
\newcommand   {\Tth}     {T_\theta}
\newcommand   {\bv}      {{\bf b}}
\newcommand   {\rv}      {{\bf r}}

\input{psfig}
\begin{document}
\begin{titlepage}

\begin{flushright}
LU TP 98-27\\
Revised version\\
May 11, 1999
\end{flushright}

\vspace{0.8in}

\LARGE
\begin{center}
{\bf Monte Carlo Study of the Phase 
Structure of Compact Polymer Chains\\} 
\vspace{.3in}
\large
Anders Irb\"ack\footnote{irback@thep.lu.se} and 
Erik Sandelin\footnote{erik@thep.lu.se}\\   
\vspace{0.10in}
Complex Systems Group, Department of Theoretical Physics\\ 
University of Lund,  S\"olvegatan 14A,  S-223 62 Lund, Sweden \\
{\tt http://www.thep.lu.se/tf2/complex/}\\
\vspace{0.3in}	

Submitted to {\it Journal of Chemical Physics}

\end{center}
\vspace{0.3in}
\normalsize
Abstract:

We study the phase behavior of single homopolymers in a
simple hydrophobic/hydrophilic off-lattice model with 
sequence independent local interactions. The specific 
heat is, not unexpectedly, found to exhibit a pronounced peak 
well below the collapse temperature, signalling a possible
low-temperature phase transition. The system size dependence 
at this maximum is investigated both with and without 
the local interactions, using chains with up to 50 monomers. 
The size dependence is found to be weak. 
The specific heat itself seems not to diverge. 
The homopolymer results are compared with those 
for two non-uniform sequences. Our calculations are performed 
using the methods of simulated and parallel tempering. The 
performances of these algorithms are discussed, 
based on careful tests for a small system. 

\end{titlepage}

\section{Introduction}

The thermodynamic behavior of isolated homopolymers is known in quite 
some detail at and above their collapse temperature $T=\Tth$, from 
analytical work~\cite{deGennes:88,deGennes:75,Grosberg:92} and numerical 
simulations of very long chains~\cite{Grassberger:95,Grassberger:97}. 
Much less is known about the behavior at low temperatures. Consequently, 
it is of utmost interest to examine the low-$T$ phase behavior  
and its model dependence, which, in particular, may shed light 
on the mechanism of folding for heteropolymers.  
  
The possibility of an interesting low-$T$ phase structure for 
homopolymers was predicted several years ago~\cite{Lifshitz:78}. 
Recent studies~\cite{Kolinski:86,Doniach:96,Bastolla:97,Doye:98}  
of lattice chains with stiffness~\cite{Flory:56} have shown 
that such chains, in addition to the coil-globule transition 
at $\Tth$, exhibit a phase transition from the 
globule phase to a frozen crystalline phase. 
Also, Zhou \etal~\cite{Zhou:96,Zhou:97} have found evidence 
for two or more phase transitions in simple off-lattice 
models without stiffness terms. 

In this paper, our starting point is a simple 
off-lattice model~\cite{Irback:97} for protein folding with two types 
of monomers, A (hydrophobic) and B (hydrophilic). In addition to 
``hydrophobicity'' forces, the model also contains sequence independent 
local interactions. It has previously been demonstrated~\cite{Irback:97}  
that it is possible to find AB sequences 
with compact, stable native structures in this model.
Furthermore, it was shown that this property is crucially dependent 
on the local interactions. In the present paper we perform a 
detailed study of A homopolymers in this model, both with 
and without the local interactions. The most striking feature of
the A homopolymers, at least for small chain lengths, is a pronounced 
peak in the specific heat, located well below $\Tth$. This could
signal that these chains, like lattice chains with stiffness,  
undergo a low-$T$ phase transition. It is therefore interesting
to look into the system size dependence at these temperatures,
which is a computationally demanding task.   

In this paper we study the behavior at the low-$T$ maximum of the specific
heat for chains with up to $N=50$ monomers, by using the methods of 
simulated~\cite{Lyubartsev:92,Marinari:92,Irback:95b} and 
parallel~\cite{Geyer:93,Tesi:96,Hukushima:96,Hansmann:97a} tempering,
which are found to be much more efficient than standard methods (see below). 
In this range of $N$, we find that the size dependence is very weak. 
There is no sign that the specific heat, or any other 
quantity, develops a singularity with increasing $N$. This holds both with
and without the local interactions, although 
the crossover is somewhat more abrupt in the presence 
of these interactions. 

We also compare homopolymer results to those for two non-uniform
AB sequences; three $N=20$ sequences, the A homopolymer and two 
AB heteropolymers, are studied with and without local interactions. 
We find that the specific heat and entropy display a relatively weak 
sequence dependence over wide ranges of low $T$. 
The dependence of these thermodynamic quantities upon the local 
interactions is, by contrast, strong.   

Our ability to simulate the low-$T$ behavior of these chains
in a controlled way relies heavily on the efficiency of simulated
and parallel tempering. In order to assess the efficiencies of these 
algorithms, we have performed careful tests for a chain with 
twelve monomers. We find that the two methods are very similar 
in efficiency, and that they are much better than a standard Monte Carlo. 

The paper is organized as follows. Section~\ref{sec_methods} contains
a brief description of the model and the two Monte Carlo algorithms, and a 
discussion of the algorithm tests. In Sec.~\ref{sec_results} we present 
the results of our simulations. A summary is given 
in Sec.~\ref{sec_discussion}.
\section{Methods}
\label{sec_methods}
\subsection{The Model}
\label{sub_model}
The model studied~\cite{Irback:97} has two types of monomers, 
A (hydrophobic) and B (hydrophilic). The monomers are linked by
rigid bonds of unit length to form linear chains. The positions
of the monomers will be denoted by $\rv_i$ ($i=1,\ldots,N$), 
and the (unit) bond vectors by $\bv_i=\rv_{i+1}-\rv_i$ ($i=1,\ldots,N-1$). 
The energy function is given by 
\beq
E=\kappa_1E_1+\kappa_2E_2+\ELJ\,,
\label{e}\eeq
where
\bea
&&E_1=-\sum_{i=1}^{N-2}\bv_i\cdot\bv_{i+1}\,,\label{e1}\\  
&&E_2=-\sum_{i=1}^{N-3}\bv_i\cdot\bv_{i+2}\,,\label{e2}\\ 
&&\ELJ=4\sum_{i=1}^{N-2}\sum_{j=i+2}^N\epsilon_{ij}\left(\frac{1}{r_{ij}^{12}}-
\frac{1}{r_{ij}^{6}}\right)\,.\label{elj}
\eea   
The model has three parameters, $\kappa_1$, $\kappa_2$ and $\epsilon_{ij}$.
The parameter $\epsilon_{ij}$ sets the depth of the minimum of the 
nonlocal Lennard-Jones interactions. It is 1 for an AA pair 
and $1/2$ for AB and BB pairs. The parameters $\kappa_1$ and  
$\kappa_2$ determine the strengths of the sequence independent 
local interactions. We will study the two choices
$(\kappa_1,\kappa_2)=(0,0)$ and $(\kappa_1,\kappa_2)=(-1,0.5)$. 
It has previously been shown~\cite{Irback:97} that there are AB sequences 
with stable native structures at relatively high temperatures for 
$(\kappa_1,\kappa_2)=(-1,0.5)$, whereas there are no, or at least 
extremely few, such sequences for $(\kappa_1,\kappa_2)=(0,0)$.

In this paper the main focus is on A homopolymers. In addition, 
we study two AB heteropolymers, denoted by S and U, which can be found 
in Table~\ref{tab:1}. The folding temperature, defined as the temperature 
where the dominance of the ground state sets in, is high for sequence S, 
and more typical for sequence U, for $(\kappa_1,\kappa_2)=(-1,0.5)$.

\begin{table}
\begin{center}
\begin{tabular}{ccc}
&Sequence & $T_f$\\
\hline
S & AAAA BBAA AABA ABAA ABBA & 0.23\\
U & BAAA AAAB AAAA BAAB AABB & $<0.15$\\
\hline
\end{tabular}
\caption{Two AB sequences and their folding temperatures 
$T_f$ for $(\kappa_1,\kappa_2)=(-1,0.5)$~\protect\cite{Irback:97}.}
\label{tab:1}
\end{center}
\end{table}           

Our $(\kappa_1,\kappa_2)=(0,0)$ A homopolymers are almost identical  
to the chains studied by Baumg\"artner~\cite{Baumgartner:80a,Baumgartner:80b} 
several years ago, except that the minimum of our Lennard-Jones potential 
is at $r_{ij}=2^{1/6}$ rather than 1. The model studied by 
Zhou~\etal~\cite{Zhou:96,Zhou:97}, for fixed and flexible bond lengths, 
has a square-well potential in place of the Lennard-Jones potential.   
\subsection{Monte Carlo Methods}  
\label{sub_algo}
Our simulations are performed using  
simulated~\cite{Lyubartsev:92,Marinari:92,Irback:95b} and 
parallel~\cite{Geyer:93,Tesi:96,Hukushima:96,Hansmann:97a} tempering. 
The basic idea of both these methods is to create an enlarged 
configuration space. This means that a number of different 
temperatures are studied simultaneously, which in particular    
can be used as a tool for speeding up the evolution of the 
system at low temperatures.

In simulated tempering one simulates the joint probability
distribution 
\beq
P(r,k)\propto \exp\left[-g_k-E(r)/T_k\right]\,,  
\label{simtemp}\eeq
where $r=\{\rv_1,\ldots,\rv_N\}$ denotes the chain conformation
and $k$ is a temperature index. This distribution contains two
sets of parameters; the temperatures that the system 
is allowed to visit, $\{T_k\}$, and a set of tunable 
simulation parameters, $\{g_k\}$. The parameters $g_k$ are free 
in the sense that averages corresponding to the desired Boltzmann 
distribution can be obtained directly, without reweighting, 
independently of these. However, the $g_k$ parameters 
govern the marginal distribution in $k$, which can be written as 
\beq
P(k)\propto\exp(-g_k-F_k/T_k)\,,
\label{marginal}\eeq
where $F_k$ is the free energy at $T=T_k$. Hence, to have good
mobility in $k$, it is essential to make a careful choice of $\{g_k\}$.   
This is typically achieved by means of trial runs. The actual 
simulation of the distribution $P(r,k)$ can be done by using 
ordinary separate updates of $r$ and $k$.

Parallel tempering uses an even larger configuration space; the 
simulated distribution is 
\beq
P(r_1,\ldots,r_K)\propto\prod_{k=1}^K\exp[-E(r_k)/T_k]\,,
\label{partemp}
\eeq
where each $r_k$ represents a complete chain conformation.  
The simulation is carried out by using two types of updates:
ordinary, parallel updates of the different $r_k$ and  
accept/reject controlled swaps $r_k\leftrightarrow r_{k+1}$.
The latter update is the analog of the $k$ update in 
simulated tempering. Compared to simulated tempering, 
this algorithm has two advantages. First, it is very 
easy to parallelize; and second, there are no $g_k$ parameters 
to be tuned. These parameters are not needed since the 
$k$ distribution is uniform by construction.  

Using these two methods, we have studied chains with 12--400 
monomers. Our simulations for $N\le 32$ were carried out 
on workstations using simulated tempering, whereas those
for larger $N$ were done on a parallel computer using parallel
tempering. The temperatures $T_k$ were chosen 
according to~\cite{Hansmann:97b}
\beq
T_k=T_{\min}\left({T_{\max}\over T_{\min}}\right)^{(k-1)/(K-1)}\,,     
\label{Tk}\eeq
where $T_{\min}=T_1$ and $T_{\max}=T_K$ denote the lowest and
highest allowed temperatures, respectively. The parameters 
$T_{\min}$, $T_{\max}$ and $K$ can be found in Table~\ref{tab:2}.

\begin{table}
\begin{center}
\begin{tabular}{ccccccc}
\cline{2-7}
\multicolumn{1}{c}{}
&\multicolumn{3}{c}{$(\kappa_1,\kappa_2)=(0,0)$}
&\multicolumn{3}{c}{$(\kappa_1,\kappa_2)=(-1,0.5)$}\\
\hline
$N$ & $T_{\min}$ & $T_{\max}$ & $K$ & $T_{\min}$ & $T_{\max}$ & $K$ \\
\hline
12 & 0.15 & 4 & 20 & 0.15 & 4 & 20 \\
20 & 0.15 & 4 & 25 & 0.15 & 4 & 25 \\
32 & 0.25 & 4 & 25 & 0.25 & 4 & 25 \\
50 & 0.25 & 4 & 35 & 0.35 & 4 & 30  \\
100 & 1.0 & 5 & 40 &  &  & \\
200 & 1.5 & 4.5 & 35 &   &   &   \\
400 & 2.0 & 5 & 30 &   &  &   \\
\hline
\end{tabular}
\caption{The parameters $T_{\min}$, $T_{\max}$ and $K$ 
[see Eq.~(\protect\ref{Tk})].}
\label{tab:2}
\end{center}
\end{table}           

\subsection{The Dynamics of the Algorithms}
\label{sub_algotest}

In order to compare the performances of simulated and parallel tempering,
we carried out calculations with both methods for the $N=12$ A homopolymer 
with $(\kappa_1,\kappa_2)=(-1,0.5)$. In these calculations we used the same 
set of temperatures (see Table~\ref{tab:2}), and the same type of 
conformation update. We also performed a standard fixed-$T$ 
simulation at $T=T_{\min}=0.15$, using the same conformation update.
  
\begin{figure}[t]
\vspace{-40mm}
\mbox{
\hspace{0mm}
\psfig{figure=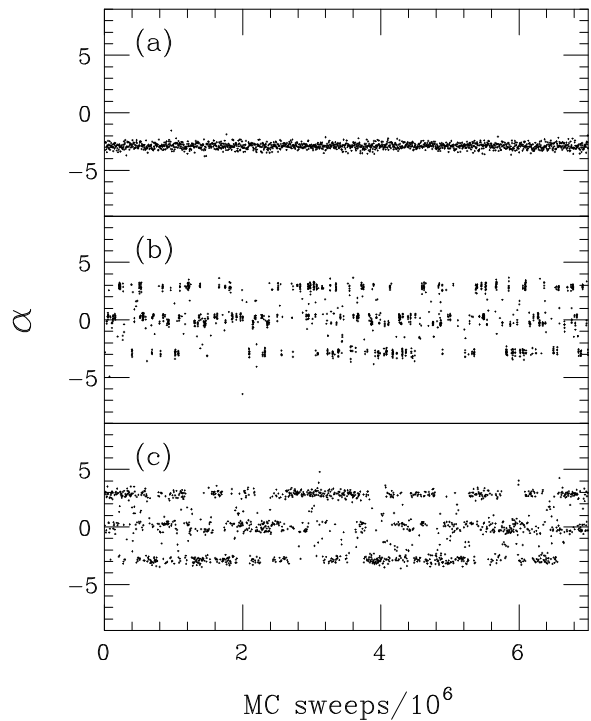,width=12cm,height=16.8cm}
}
\vspace{-25mm}
\caption{
Evolution of the sum of all torsional angles, $\alpha$, 
in three different simulations of the $N=12$ A homopolymer for 
$(\kappa_1,\kappa_2)=(-1,0.5)$. (a) Standard Monte Carlo, (b) simulated 
tempering, and (c) parallel tempering. All data shown correspond to 
$T=T_{\min}=0.15$. 
} 
\label{fig:1}
\end{figure}

In Fig.~\ref{fig:1} we show run-time histories for the sum of 
all torsional angles along the chain, $\alpha$, from these  
three simulations. The quantity $\alpha$ is convenient to study 
since it is known from the symmetry of the model that its probability 
distribution has to be symmetric, $P(\alpha)=P(-\alpha)$. When 
comparing the different runs in Fig.~\ref{fig:1}, the horizontal axis 
can be thought of as CPU time, since the cost of the additional 
updates used in simulated and parallel tempering is negligible.          
Therefore, it is evident from this figure that both simulated 
and parallel tempering are dramatic improvements on the 
standard method. Note that the system is trapped at negative 
$\alpha$ during the whole fixed-$T$ simulation, whereas the true 
distribution $P(\alpha)$ is symmetric.

Deciding which is best of simulated and parallel tempering is    
a more delicate task, which requires a careful analysis of the
statistical errors. To get accurate estimates of the errors, 
we performed two additional runs, a factor of 10 longer than those 
in Fig.~\ref{fig:1}. The data from these long runs were analyzed
by using the jackknife method~\cite{Miller:74}, which is
chosen because the number of visits of a given temperature is a 
random variable in simulated tempering. As a result,    
one does not have an unbiased estimator of, say, $\alpha$ 
at temperature $T=T_k$, but rather one estimates the ratio
\beq
\ev{\alpha}_{T_k}={\ev{\chi_k\cdot\alpha}\over\ev{\chi_k}}\,,  
\eeq
where $\ev{\cdot}$ without subscript denotes an average with 
respect to the joint distribution [see Eq.~(\ref{simtemp})], 
and $\chi_k$ is 1 if $T=T_k$ and 0 otherwise. For parallel
tempering the jackknife procedure is equivalent to a standard 
block analysis. 

The jackknife analysis was carried out for different bin sizes.
For simulated tempering the dependence on bin size was weak.
For parallel tempering, it was, by contrast, necessary to
use relatively large bin sizes in order to get stable error estimates
at low $T$. For this method, we also computed the 
autocorrelation function for $\alpha$. 
Consistent with the jackknife analysis, the exponential 
autocorrelation time was estimated to be approximately $7\times10^4$ 
sweeps (same unit as in Fig.~\ref{fig:1}) at $T=T_{\min}$. 
 
The final statistical errors for the two methods differed by 
less than 10\%. Thus, the algorithms are very similar in 
efficiency for this system. This is in line with the findings 
of Hansmann and Okamoto~\cite{Hansmann:97b}, who measured 
the time needed to find the ground state in an all-atom model 
for a small peptide, with similar results for simulated 
and parallel tempering.    
\subsection{Measurements}
\label{sub_meas}
Two important observables in our study of the low-$T$ behavior 
of these chains are the specific heat
\beq
C_v={\partial\ev{E}\over \partial T}=
{1\over T^2}\left(\ev{E^2}-\ev{E}^2\right)\, 
\label{C}\eeq
and the bond-bond correlation function
\beq
C_b(d)={1\over N-d}\sum_{i=1}^{N-d}\bv_i\cdot\bv_{i+d}\,,
\label{Cb}\eeq
which for $d=1$ and 2 can be written as $C_b(d)=-E_d/(N-d)$ 
[see Eqs.~(\ref{e1}) and~(\ref{e2})].      

We estimate the collapse temperature $\Tth$ by using the 
radius of gyration $r_g$, defined by 
\beq
r_g^2={1\over N}\sum_{i=1}^N \left(\rv_i - 
\frac{1}{N}\sum_{j=1}^N\rv_j\right)^2\,. 
\label{cv}\eeq
The ratio $\ev{r_g^2}/N$ tends to zero for fixed $T<\Tth$, 
and to infinity for fixed $T>\Tth$, as $N\to\infty$. As is well-known, 
this means that $\Tth$ can be estimated from the   
intersections of curves corresponding to different $N$, in
a plot of $\ev{r_g^2}/N$ vs $T$.

In our comparison of different $N=20$ chains, we calculate their   
entropies $S$. The simulated-tempering algorithm is well suited 
for this purpose, since free-energy differences can be 
directly obtained from the marginal probabilities 
$P(k)$~\cite{Lyubartsev:92}. Using Eq.~(\ref{marginal}), one 
finds that
\beq
S_{k}-S_{l}=g_k-g_l+\ln{P(k)\over P(l)}+
{\ev{E}_k\over T_k}-{\ev{E}_l\over T_l}\,,     
\label{entropy}\eeq
where $S_{k,l}$ and $\ev{E}_{k,l}$ denote the entropy and 
average energy, respectively, at $T=T_{k,l}$. This expression is handy 
since the parameters $g_k$ are constants and the average energies are 
relatively easy to measure. The remaining term, $\ln P(k)/P(l)$,
is just a small correction if the parameters $g_k$ are well chosen.   

Equation~(\ref{entropy}) can be used to obtain the temperature dependence 
for a given chain. In order to compare the entropies of two 
different chains [different sequences and/or different $(\kappa_1,\kappa_2)$,
but the same $N$], we use the umbrellalike~\cite{Torrie:77} formula  
\beq
S^\prime-S={\ev{E^\prime}^\prime-\ev{E}\over T}
+\ln\,\left\langle\exp\left(-{E^\prime-E\over T}\right)\right\rangle\,,
\label{entropy2}\eeq     
where the prime is used to distinguish the two systems. In 
Eq.~(\ref{entropy2}) 
one could, of course, replace $\ev{E^\prime}^\prime$ by expectation 
values referring to the system without prime.  

All statistical errors quoted below are 1$\sigma$ errors, obtained
by the jackknife method~\cite{Miller:74}.   

\section{Results}
\label{sec_results}
\begin{figure}[t]
\vspace{-40mm}
\mbox{
  \hspace{-35mm}
  \psfig{figure=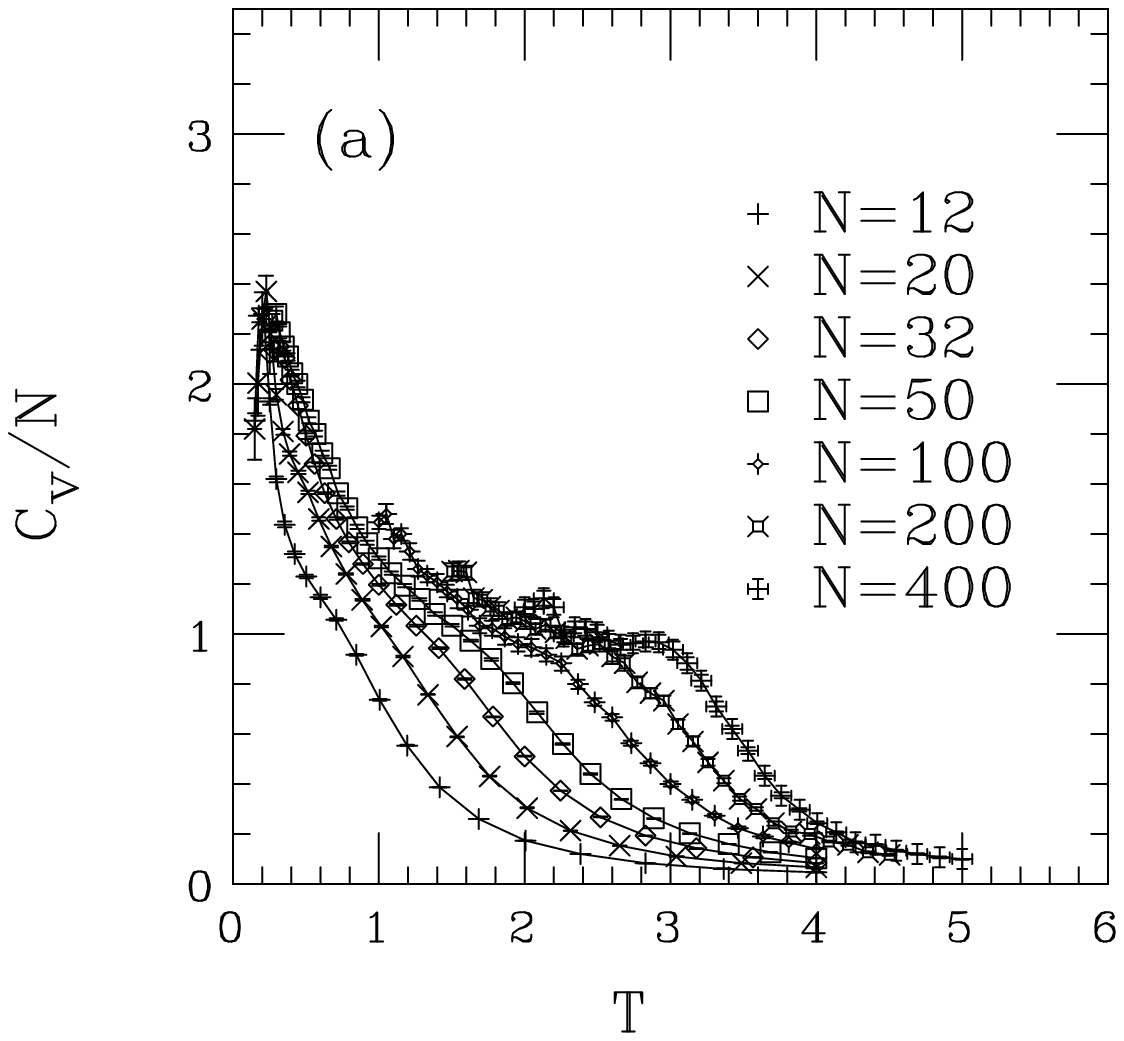,width=10.5cm,height=14cm}
  \hspace{-35mm}
  \psfig{figure=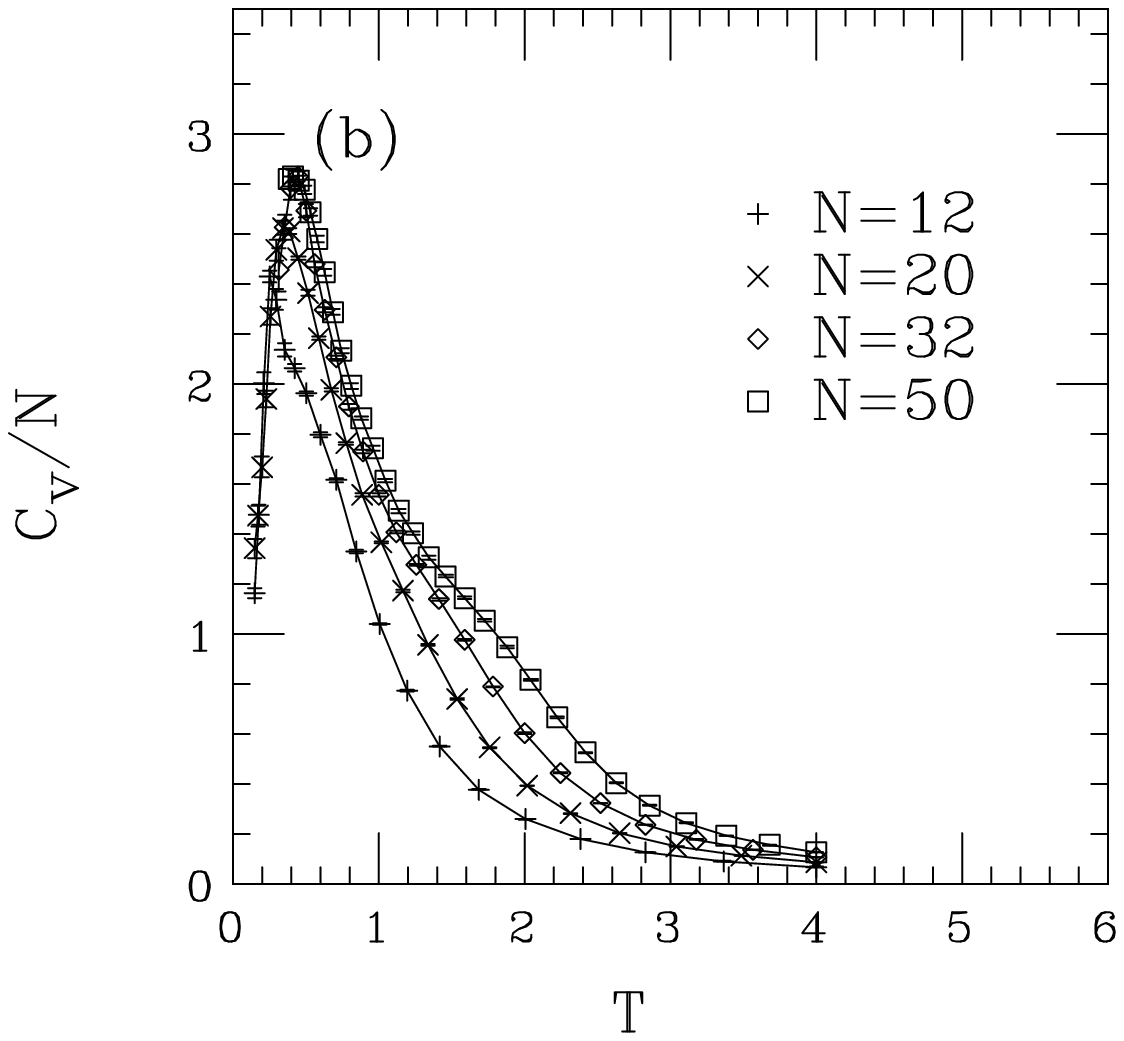,width=10.5cm,height=14cm}
}
\vspace{-40mm}
\caption{Temperature dependence of the specific heat for A homopolymers  
for (a) $(\kappa_1,\kappa_2)=(0,0)$ and (b)
$(\kappa_1,\kappa_2)=(-1,0.5)$.}
\label{fig:2}
\end{figure}

We have performed simulations of the A homopolymers of the model 
defined in Sec.~\ref{sub_model} for $(\kappa_1,\kappa_2)=(0,0)$ 
and $(-1,0.5)$. In Fig.~\ref{fig:2} we show the temperature dependence
of the specific heat for different chain lengths. For both choices 
of $(\kappa_1,\kappa_2)$, it can be seen that the specific heat 
exhibits a pronounced peak at low temperatures, which could 
signal a phase transition. The height of the peak depends, however,  
only weakly on $N$. The $N$ dependence is stronger at higher 
temperatures where another peak seems to develop. This shoulder or 
peak can most probably be associated with the collapse transition 
at $\Tth$. In Sec.~\ref{sub_theta} we estimate $\Tth$ by using 
data for the radius of gyration. We then return to the $N$ 
dependence at low temperatures in Sec.~\ref{sub_fss}. Finally, in 
Sec.~\ref{sub_mixed}, we compare the behaviors of three different  
$N=20$ sequences, the A homopolymer and the two AB heteropolymers in 
Table~\ref{tab:1}.      

\subsection{Estimation of $\Tth$}
\label{sub_theta}

\begin{figure}[t]
\vspace{-40mm}
\mbox{
\hspace{-35mm}
\psfig{figure=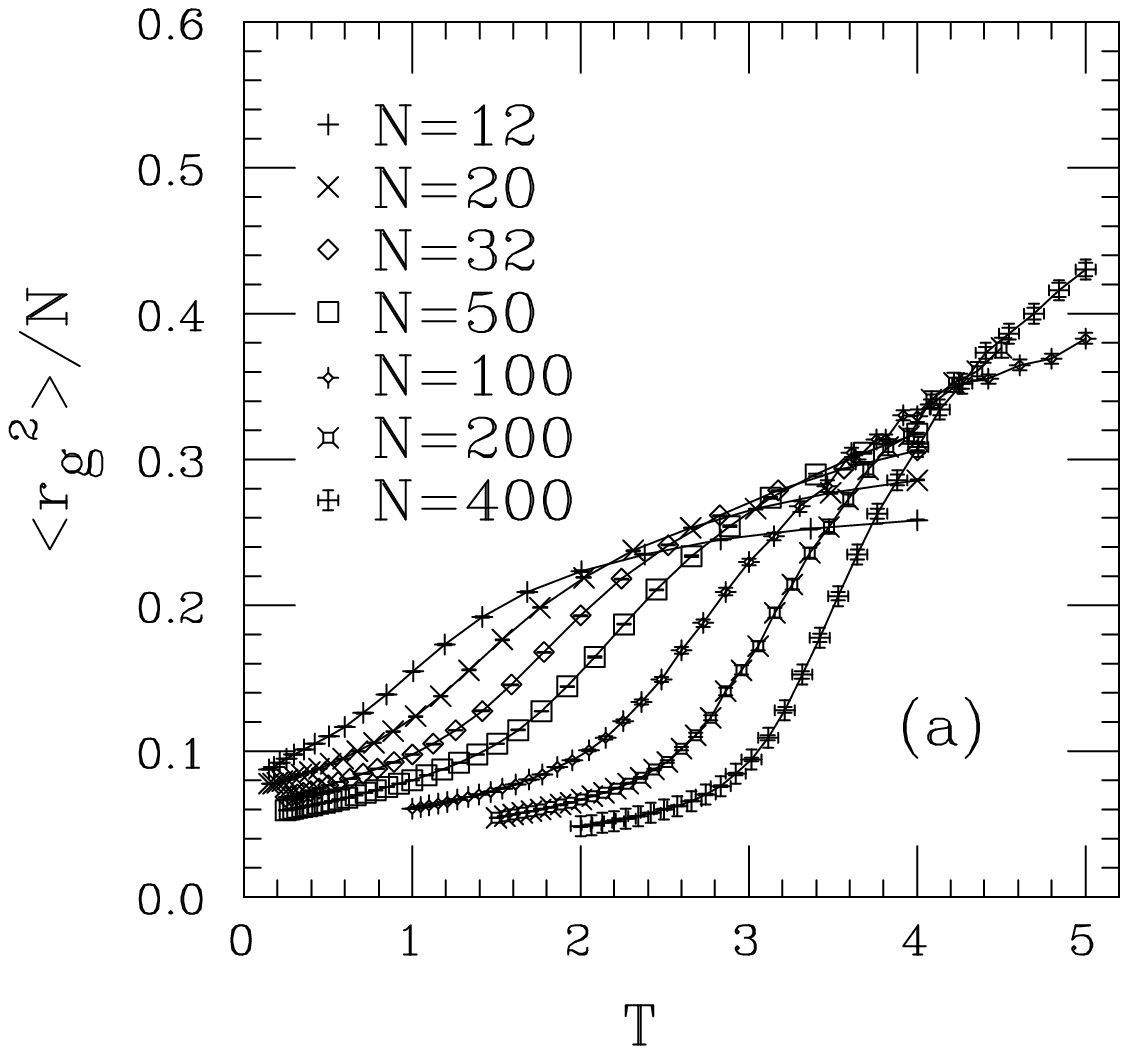,width=10.5cm,height=14cm}
\hspace{-35mm}
\psfig{figure=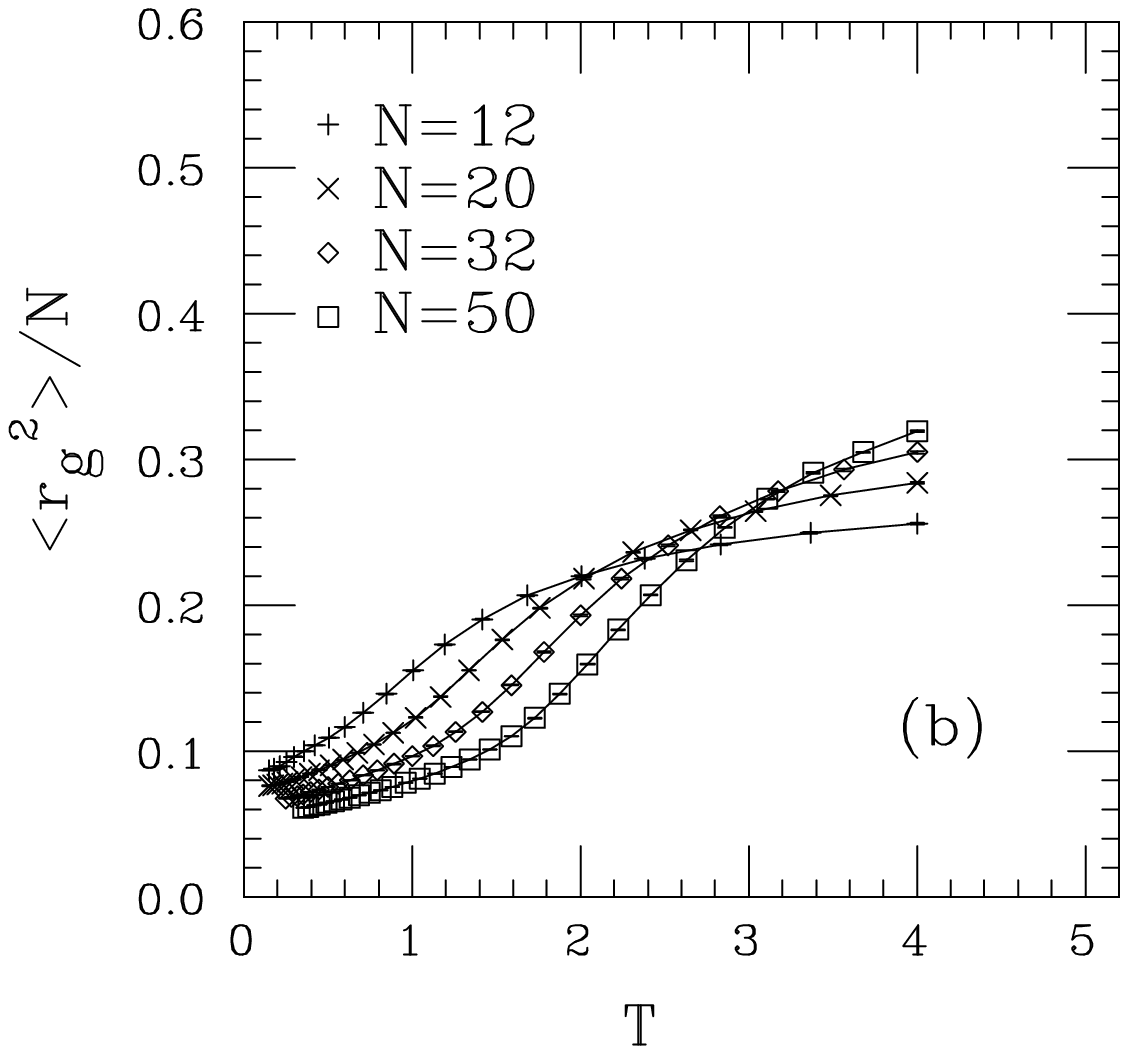,width=10.5cm,height=14cm}
}
\vspace{-40mm}
\caption{Temperature dependence of $\ev{\rgyr^2}/N$ for A homopolymers
for (a) $(\kappa_1,\kappa_2)=(0,0)$ and (b) $(\kappa_1,\kappa_2)=(-1,0.5)$.}
\label{fig:3}
\end{figure}

In order to estimate $\Tth$, it is easier to use the
radius of gyration than the specific heat. In Fig.~\ref{fig:3}a we show 
the ratio $\ev{r_g^2}/N$ against $T$ for different $N$ 
for $(\kappa_1,\kappa_2)=(0,0)$. In the temperature range 
relevant for the $\Tth$ estimation, we studied chains
with up to 400 monomers (simulations are much easier here than
at low $T$). The temperature at which the curves corresponding 
to the sizes $N$ and $2N$ intersect, $T_N$, was estimated to be 
$3.76\pm0.11$, $4.05\pm0.06$ 
and $4.33\pm0.08$ for $N=50$, 100 and 200, respectively.     
A least-square fit of these data points to the mean-field form
\beq
T_N=\Tth+\frac{{\rm const.}}{\sqrt{N}}
\eeq
yields $\Tth=4.88\pm0.18$. The systematic uncertainty in this extrapolation
is difficult to estimate without data for larger $N$. Nevertheless, this 
analysis convincingly shows that $\Tth$ is indeed located well above the 
low-$T$ maximum of the specific heat. 

In Fig.~\ref{fig:3}b we show the same plot for $(\kappa_1,\kappa_2)=(-1,0.5)$.
All data are for $N\le50$. They are very similar to those obtained 
for $(\kappa_1,\kappa_2)=(0,0)$ at these $N$. In particular, it is 
clear that the collapse transition occurs well above
the low-$T$ maximum of the specific heat in this case too.      

\subsection{The Low-$T$ Behavior}
\label{sub_fss}

\begin{figure}[t]
\vspace{-40mm}
\mbox{
  \hspace{-35mm}
  \psfig{figure=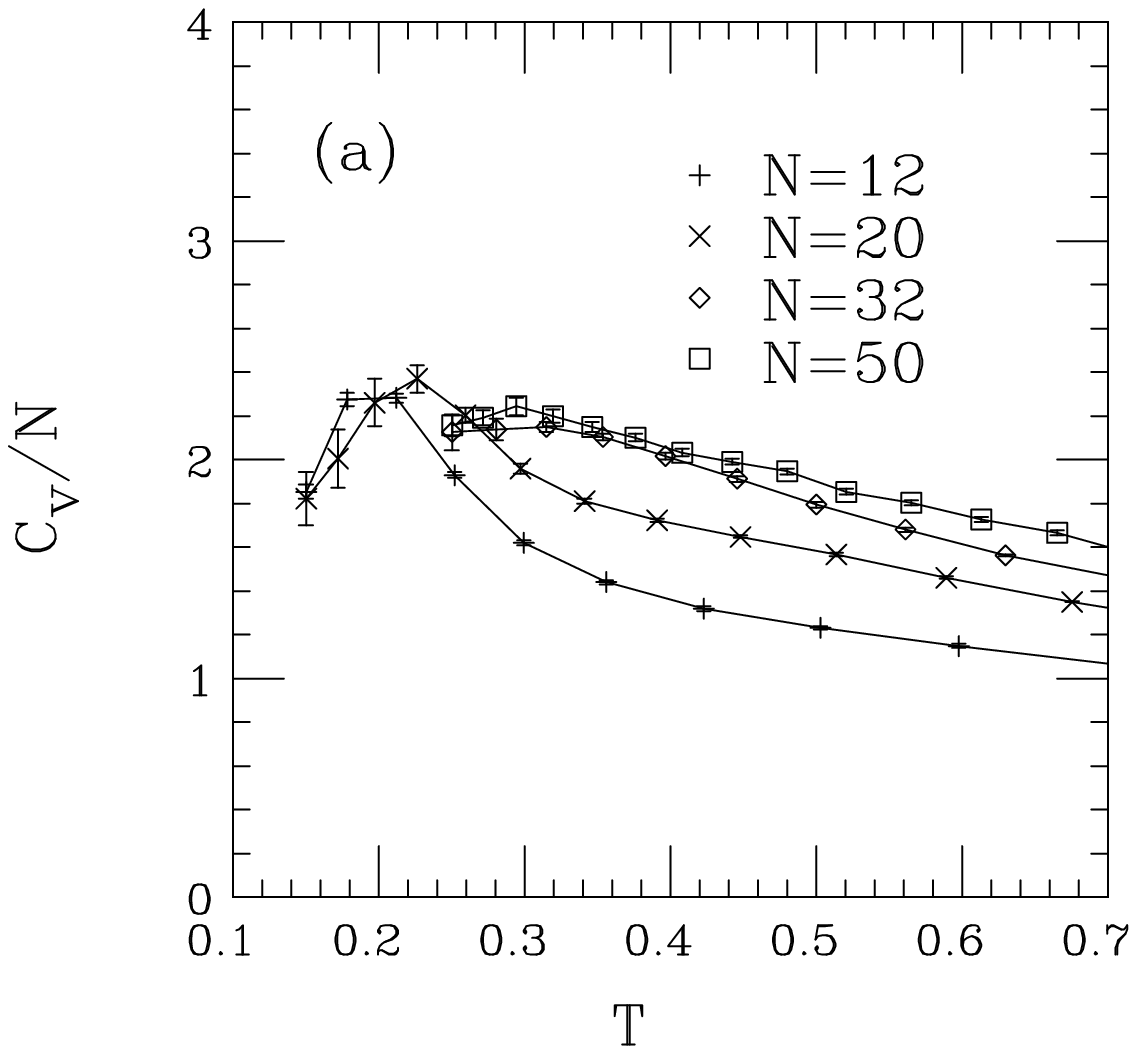,width=10.5cm,height=14cm}
  \hspace{-35mm}
  \psfig{figure=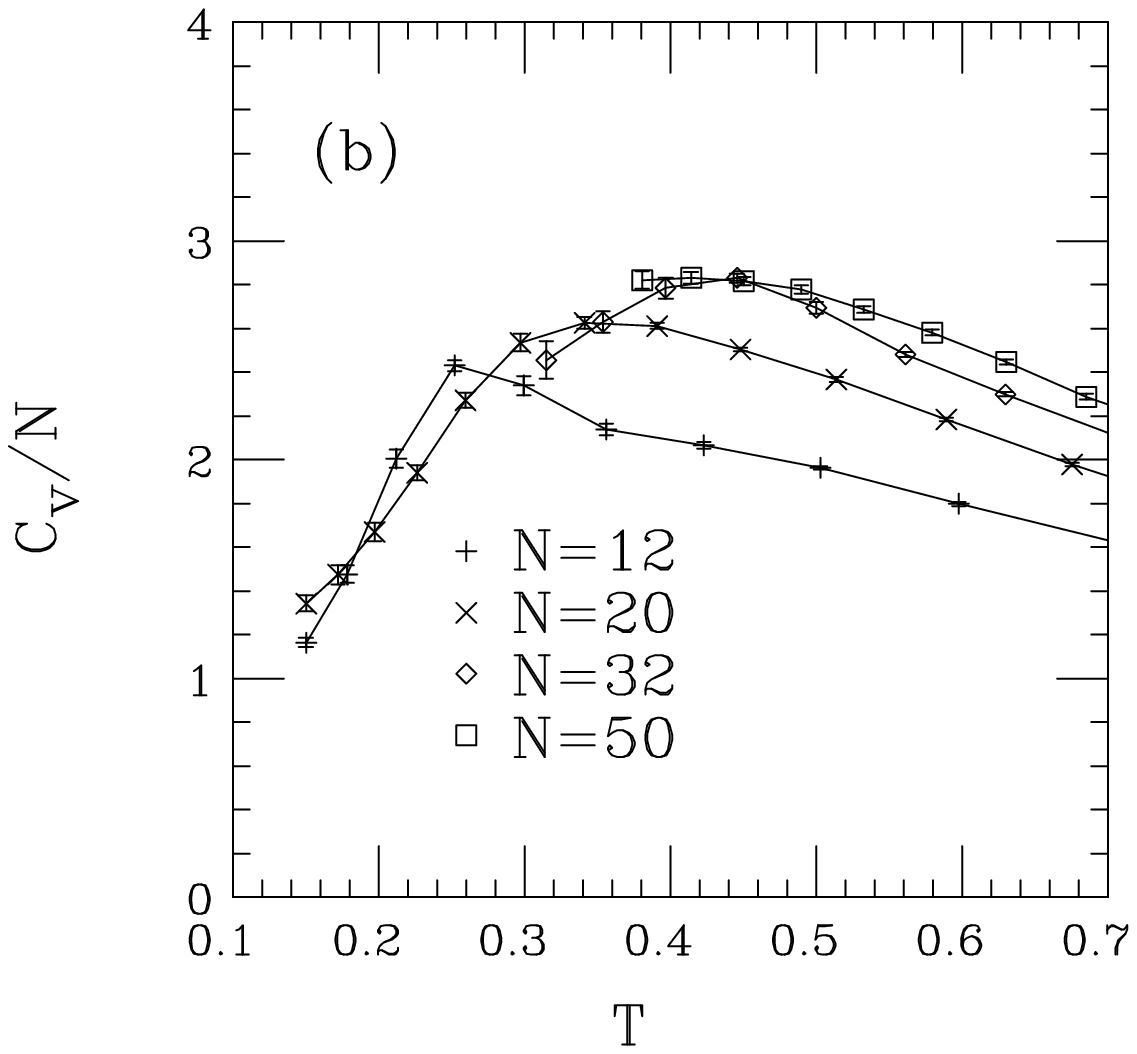,width=10.5cm,height=14cm}
}
\vspace{-40mm}
\caption{Low-$T$ behavior of the specific heat for A
homopolymers for (a) $(\kappa_1,\kappa_2)=(0,0)$ and (b)  
$(\kappa_1,\kappa_2)=(-1,0.5)$.}
\label{fig:4}
\end{figure}

We now turn to the behavior around the low-$T$ maximum of the 
specific heat. Figure~\ref{fig:4}, which is an enlargement of 
Fig.~\ref{fig:2}, shows the specific heat itself in the vicinity of the 
maximum. A linear scaling with $N$ of the height of the peak in $C_v/N$ 
would have implied a first-order phase transition. The data show 
a weak $N$ dependence. The peak grows very slowly with $N$  
for $(\kappa_1,\kappa_2)=(-1,0.5)$, and not at all 
for $(\kappa_1,\kappa_2)=(0,0)$. In neither case is there any   
sign that $C_v/N$ diverges. We also studied the energy  
distributions at the maxima. Consistent with the specific heat 
analysis, these were found to have a single peak, without 
any trace of bimodality.   

\begin{figure}[t]
\vspace{-40mm}
\mbox{
  \hspace{-35mm}
  \psfig{figure=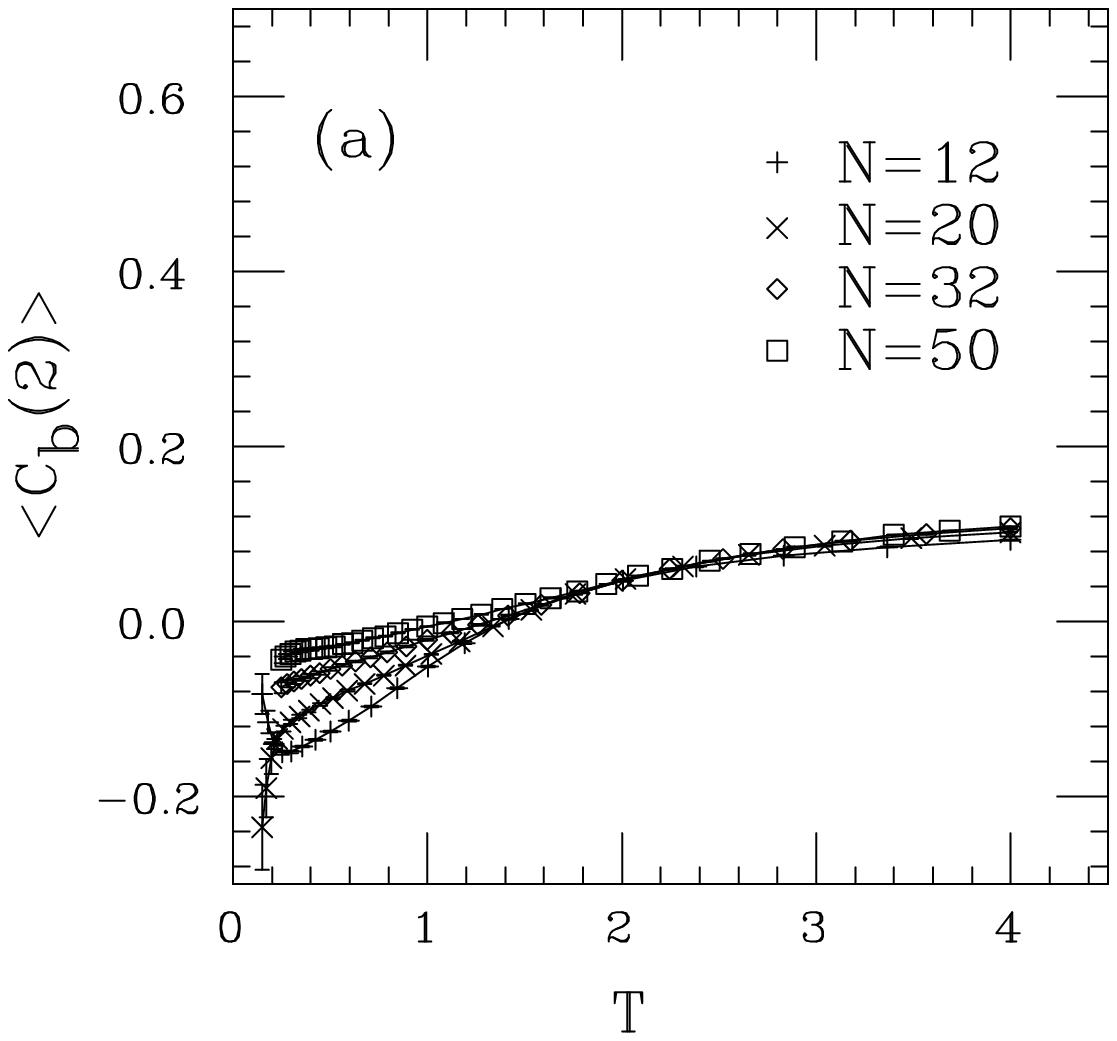,width=10.5cm,height=14cm}
  \hspace{-35mm}
  \psfig{figure=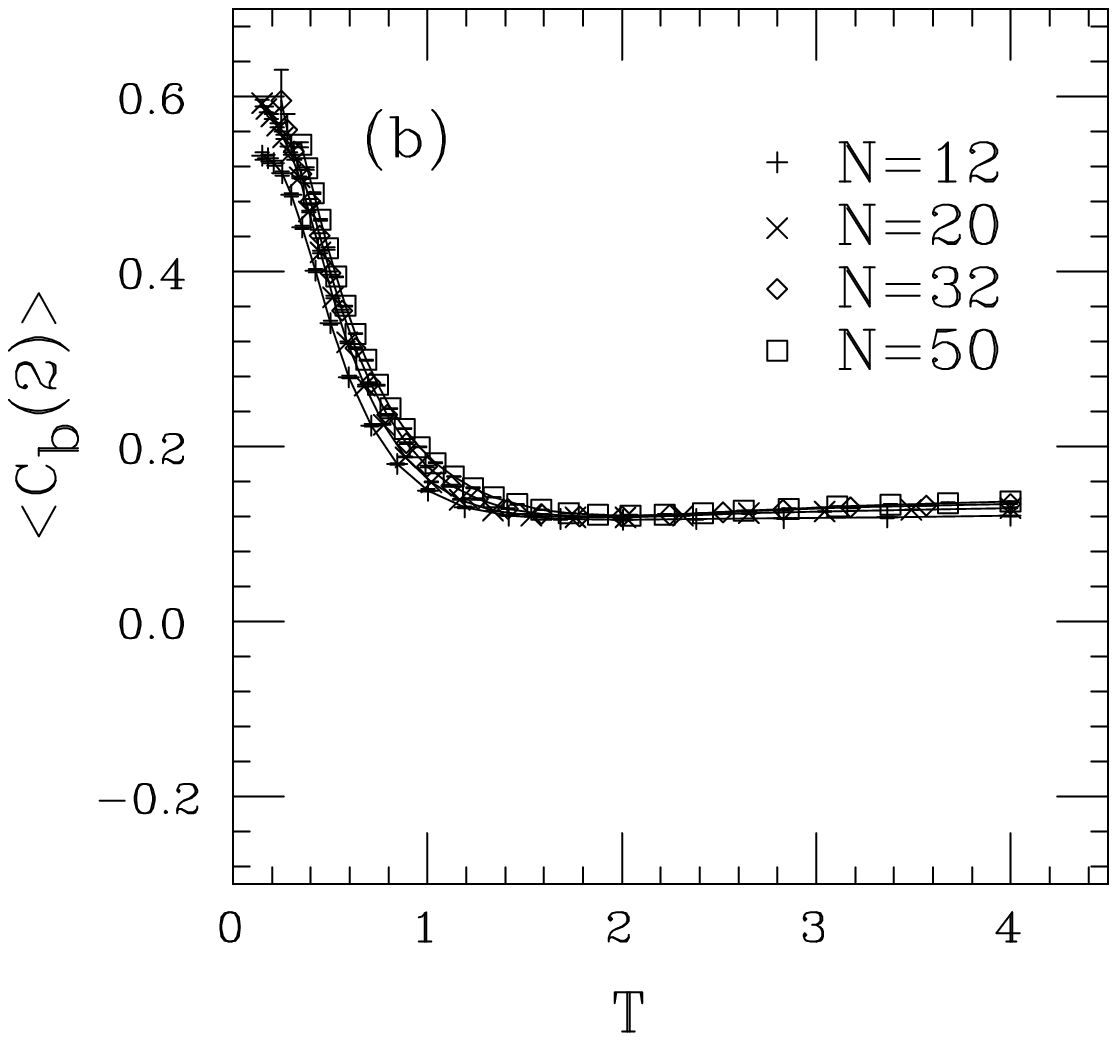,width=10.5cm,height=14cm}
}
\vspace{-40mm}
\caption{Temperature dependence of $C_b(2)$ for A homopolymers
for (a) $(\kappa_1,\kappa_2)=(0,0)$ and (b)
$(\kappa_1,\kappa_2)=(-1,0.5)$.} 
\label{fig:5}
\end{figure}

The size of the chains, as measured by the radius of gyration, 
changes very little around the low-$T$ maximum of the specific heat
(see Fig.~\ref{fig:3}). The size is similar for the two choices of
$(\kappa_1,\kappa_2)$, and very small at these temperatures.
The bond-bond correlation $C_b(d)$ [see Eq.~(\ref{Cb})] provides
information about the local structure of the chains. Unlike  
the radius of gyration, this quantity is strongly 
$(\kappa_1,\kappa_2)$ dependent. This can be seen from  
Fig.~\ref{fig:5}, which shows $C_b(d)$ against $T$ for $d=2$. 
The behavior of $C_b(2)$ for $(\kappa_1,\kappa_2)=(-1,0.5)$ shows that
strong regularities develop in the local structure at low $T$. 
The change with decreasing $T$ is, however, gradual, and 
does not become more abrupt with increasing $N$. In the absence of
local interactions, $(\kappa_1,\kappa_2)=(0,0)$, the
correlations $C_b(d)$ remain weak at low $T$. For small $N$ there are
some irregularities at very low $T$, but these disappear
with increasing $N$. 

Let us compare these results with findings for lattice chains.
As mentioned in the introduction, it has been 
found~\cite{Kolinski:86,Doniach:96,Bastolla:97,Doye:98}  
that lattice chains with stiffness exhibit a transition
to an ordered low-$T$ phase. Our $(\kappa_1,\kappa_2)=(-1,0.5)$ 
chains, which have a certain degree of stiffness, do show strong
regularities in the local structure at low $T$, as can be seen  
from the behavior of the correlation $C_b(d)$. Also, the peak 
in the specific heat is more pronounced for these chains than
for the $(\kappa_1,\kappa_2)=(0,0)$ chains. However, we do
not find any sign that a phase transition takes place. 
Although the $N$ range probed is comparatively small, 
this indicates a possible difference between lattice and 
off-lattice chains. It would be interesting to systematically 
study how the off-lattice results depend on the degree of 
stiffness of the chains, but this is beyond the scope of the 
present paper.

The low-$T$ behavior of off-lattice models similar to ours
has been studied before. As mentioned earlier, 
Baumg\"artner~\cite{Baumgartner:80a,Baumgartner:80b}
studied a model very closely related to our 
$(\kappa_1,\kappa_2)=(0,0)$ model. Here the low-$T$ peak in 
the specific heat was associated with a probably 
first-order phase transition. As we have seen, our results
provide no support for the existence of such a transition.
Zhou~\etal~\cite{Zhou:96,Zhou:97} studied a model with 
square-well potentials, both for fixed and flexible bond lengths.
For flexible bond lengths, evidence in terms of bimodal energy 
distributions was found for a first-order low-$T$ phase transition. 
This contrasts sharply with the behavior of our chains, which,
as mentioned above, have unimodal energy distributions. 

\subsection{Heteropolymers} 
\label{sub_mixed}

In this section we compare the thermodynamic behaviors of  
the $N=20$ A homopolymer and the two $N=20$ heteropolymers S and U 
in Table~\ref{tab:1}. Each of the three sequences is studied both 
for $(\kappa_1,\kappa_2)=(0,0)$ and $(-1,0.5)$, 
which gives us a total of six different chains. The temperatures
studied range between 0.15 and 4. One of the chains,
sequence S with $(\kappa_1,\kappa_2)=(-1,0.5)$, has a
folding temperature $T_f\approx 0.23>0.15$~\cite{Irback:97} 
and is referred to as the stable one, whereas $T_f<0.15$ 
for the others.  

\begin{figure}[t]
\vspace{-40mm}
\mbox{
  \hspace{-35mm}
  \psfig{figure=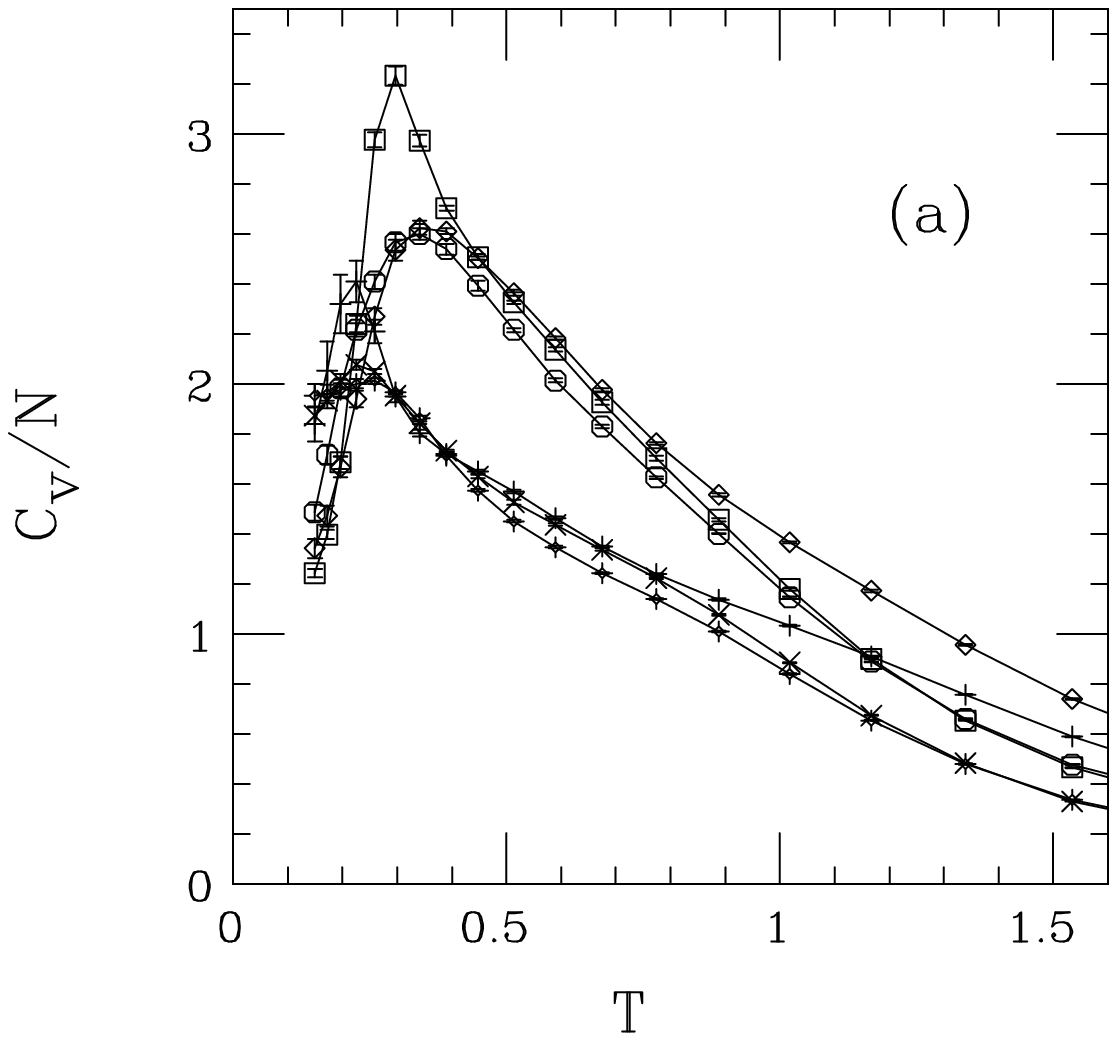,width=10.5cm,height=14cm}
  \hspace{-35mm}
  \psfig{figure=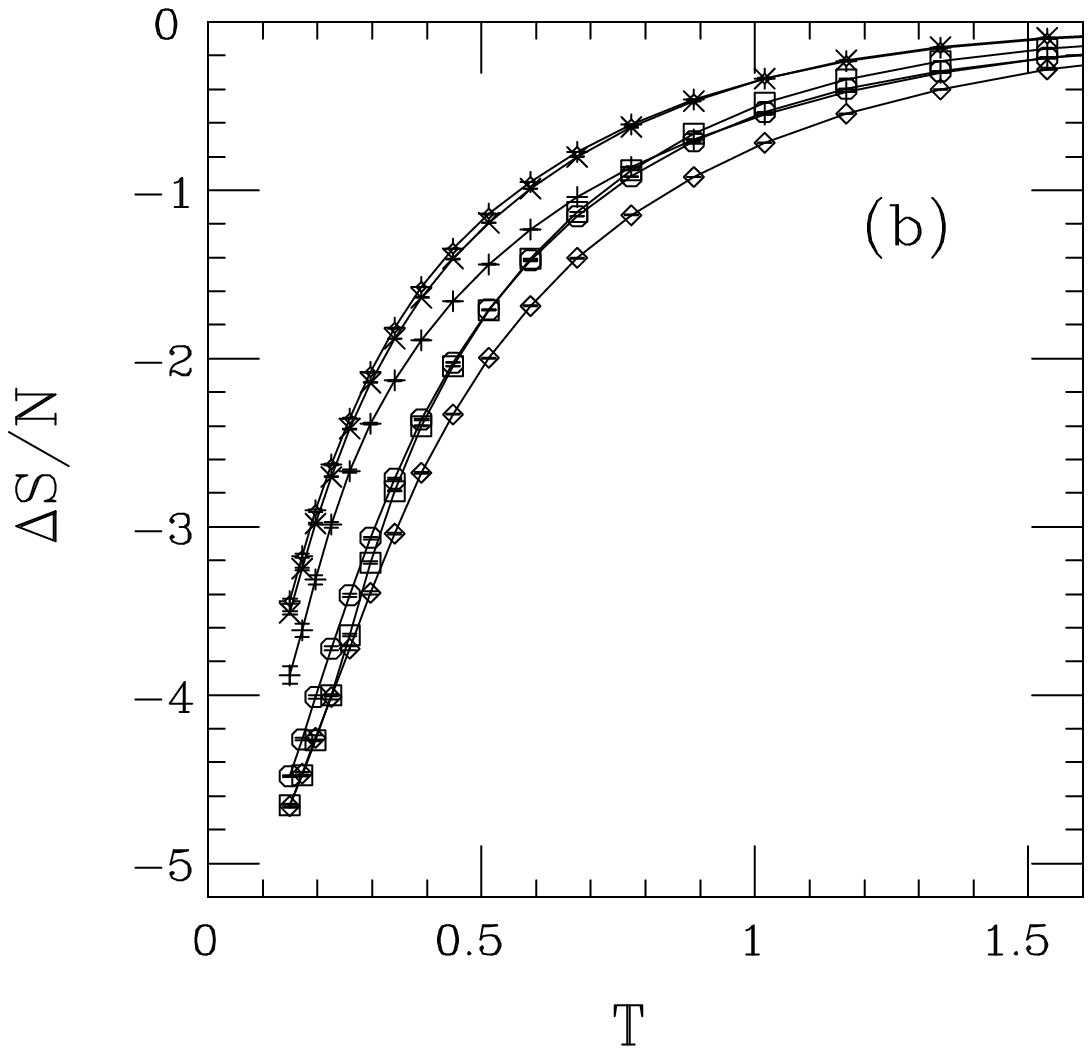,width=10.5cm,height=14cm}
}
\vspace{-40mm}
\caption{
(a) Specific heat and (b) entropy against temperature for 
the following six $N=20$ chains: 
A homopolymer, $(\kappa_1,\kappa_2)=(0,0)$ ($+$);
sequence S, $(\kappa_1,\kappa_2)=(0,0)$ ($\times$);  
sequence U, $(\kappa_1,\kappa_2)=(0,0)$ (fancy $+$);  
A homopolymer, $(\kappa_1,\kappa_2)=(-1,0.5)$ ($\diamond$);
sequence S, $(\kappa_1,\kappa_2)=(-1,0.5)$ ($\Box$); 
sequence U, $(\kappa_1,\kappa_2)=(-1,0.5)$ ($\circ$).
}
\label{fig:6}
\end{figure}

In Fig.~\ref{fig:6}a we show the specific heat, which 
has a peak at low $T$ for all the six chains. 
Note that the peak is highest 
for the stable chain, and that the maximum is at $T\approx 0.30$ 
for this chain, which is fairly close to its folding temperature. 
The perhaps most interesting part of this figure is, however, 
between $T\approx0.4$ and $T\approx0.9$. Here the data for the 
different chains approximately collapse onto two curves, each 
corresponding to a fixed $(\kappa_1,\kappa_2)$. Thus, the 
dependence on sequence is weak over this fairly wide range 
of temperature.  

We also calculated the entropies of the chains, using the method 
described in Sec.~\ref{sub_meas}, which is convenient since 
it does not require numerical integration of any observable. In 
Fig.~\ref{fig:6}b we show $\Delta S=S-S_0$, where $S_0$ is chosen 
as the entropy of the $(\kappa_1,\kappa_2)=(-1,0.5)$ A homopolymer at $T=4$.  
Thus, the constant $S_0$ is the same for all the chains. 
From the figure it can be seen that the chains with 
$(\kappa_1,\kappa_2)=(-1,0.5)$ have lower entropies than those
with $(\kappa_1,\kappa_2)=(0,0)$ at low $T$. This is not unexpected
since the $(\kappa_1,\kappa_2)=(-1,0.5)$ chains have stiffness. 
What still may seem somewhat surprising is that the stable chain, 
below its folding temperature, does not have lower entropy 
than the A homopolymer with the same $(\kappa_1,\kappa_2)$. 
This example underlines that the entropy is largely determined by
local features at low $T$. The entropy does not tell whether 
the structure is globally stable or not (which it does in lattice models). 

\section{Summary}
\label{sec_discussion}

We have studied isolated off-lattice homopolymers with 
$[(\kappa_1,\kappa_2)=(-1,0.5)]$ and without $[(\kappa_1,\kappa_2)=(0,0)]$
local interactions. In both cases the specific heat exhibits a 
peak at low $T$. The peak is most pronounced for 
$(\kappa_1,\kappa_2)=(-1,0.5)$. Also, the local structure 
of the chains develops strong regularities at low 
$T$ for this choice of $(\kappa_1,\kappa_2)$. The dependence on
system size is, however, weak at these temperatures; 
our results, obtained for $N\le 50$, show no sign 
that a singularity develops with increasing $N$. 

We have also compared homopolymer results to those for two non-uniform 
sequences, which were deliberately chosen to represent different
types of behavior. Nevertheless, the specific heat and entropy show   
a relatively weak sequence dependence over wide ranges of low $T$. The 
dependence of these thermodynamic quantities upon the local interactions 
is, by contrast, strong.

Our calculations have been performed by using simulated and
parallel tempering. Our tests show that these two methods are indeed 
much more efficient than standard methods in the difficult low-$T$ regime.  
The difference in efficiency between simulated and parallel tempering 
was found to be small.

{\it Note added in proof}. After this manuscript was completed, we 
became aware of two studies \cite{Frenkel:92,Grassberger:95b} of 
$(\kappa_1,\kappa_2)=(0,0)$ A homopolymers. Grassberger and 
Hegger~\cite{Grassberger:95b} estimated $0.2\le1/\Tth\le0.23$, 
which is in perfect agreement with our estimate.

\section*{Acknowledgements}

This work was supported by the Swedish Foundation for Strategic
Research, by the Swedish Council for Planning and Coordination
of Research (FRN), and by Paralleldatorcentrum (PDC), 
Royal Institute of Technology.

\newpage

\newcommand  {\Biomet}   {{\it Biometrika\ }}
\newcommand  {\Biopol}   {{\it Biopolymers\ }}
\newcommand  {\BC}       {{\it Biophys.\ Chem.\ }}
\newcommand  {\BJ}       {{\it Biophys.\ J.\ }}
\newcommand  {\CPL}      {{\it Chem. Phys. Lett.\ }}
\newcommand  {\COSB}     {{\it Curr.\ Opin.\ Struct.\ Biol.\ }}
\newcommand  {\EL}       {{\it Europhys.\ Lett.\ }}
\newcommand  {\JCC}      {{\it J.\ Comput.\ Chem.\ }}
\newcommand  {\JCoP}     {{\it J.\ Comput.\ Phys.\ }}
\newcommand  {\JCP}      {{\it J.\ Chem.\ Phys.\ }}
\newcommand  {\JMB}      {{\it J.\ Mol.\ Biol.\ }}
\newcommand  {\JP}       {{\it J.\ Phys.\ }}
\newcommand  {\JPC}      {{\it J.\ Phys.\ Chem.\ }}
\newcommand  {\JPCM}     {{\it J.\ Phys.: Condens. Matter\ }}
\newcommand  {\JPSJ}     {{\it J. Phys. Soc. (Jap)\ }}
\newcommand  {\JSP}      {{\it J.\ Stat.\ Phys.\ }}
\newcommand  {\Mac}      {{\it Macromolecules\ }}
\newcommand  {\MC}       {{\it Makromol.\ Chem.,\ Theory Simul.\ }}
\newcommand  {\MP}       {{\it Molec.\ Phys.\ }}
\newcommand  {\Nat}      {{\it Nature}}
\newcommand  {\NP}       {{\it Nucl.\ Phys.}}
\newcommand  {\Pro}      {{\it Proteins:\ Struct.\ Funct.\ Genet.\ }}
\newcommand  {\ProSci}   {{\it Protein\ Sci.\ }}
\newcommand  {\Pa}       {{\it Physica\ }}
\newcommand  {\PL}       {{\it Phys.\ Lett.\ }}
\newcommand  {\PNAS}     {{\it Proc.\ Natl.\ Acad.\ Sci.\ USA\ }}
\newcommand  {\PR}       {{\it Phys.\ Rev.\ }}
\newcommand  {\PRL}      {{\it Phys.\ Rev.\ Lett.\ }}
\newcommand  {\PRS}      {{\it Proc.\ Roy.\ Soc.\ }}
\newcommand  {\RMP}      {{\it Rev.\ Mod.\ Phys.\ }}
\newcommand  {\Sci}      {{\it Science\ }}
\newcommand  {\ZP}       {{\it Z.\ Physik\ }}


\end{document}